\documentclass[prd,12pt]{revtex4}
\usepackage{color,amsmath,graphicx,epstopdf}
\usepackage{verbatim}
\usepackage{subfig}

\newcommand{\blue}[1]{\textcolor{blue}{#1}}

\begin{document}
\title{An open window to neutrino mass hierarchy at maximal Dirac CP violation}
\author{Ying Zhang\footnote{E-mail:hepzhy@mail.xjtu.edu.cn.}}
\address{School of Science, Xi'an Jiaotong University, Xi'an, 710049, China}

\begin{abstract}
Non-zero leptonic CP phases in the neutrino sector are clear evidence for physics beyond the Standard Model and have many implications in particle physics and cosmology.
Some clues to maximal Dirac CP violation $\delta_{CP}=3\pi/2$ are reviewed. 
An approach to connect the Dirac CP violation phase and the neutrino mass hierarchy in terms of the measurement of CP asymmetry  is proposed.
Under appropriate requirements for baseline distance and beam energy, the problem of neutrino mass hierarchy can be revealed at maximal Dirac CP violation.
General allowed parameter spaces are investigated, within which the two unknown neutrino issues, mass hierarchy and Dirac CP violation,  can be solved simultaneously.
The abilities of addressing the mass hierarchy of current long baseline neutrino experiments are also discussed. 

\bigskip



\end{abstract}

\maketitle

In the past several decades, substantial progress has been made in the field of neutrino physics.  
Three-flavor neutrino oscillations  have been confirmed from atmosphere, solar, reactor, accelerator and other neutrino experiments. Including non-zero $\theta_{13}$, all mixing angles have also been measured precisely \cite{review1}. 
However, there are still some unsolved problems in the neutrino sector such as the mass hierarchy, leptonic CP violation (CPV), and Dirac/ Majorana fermions. These principal issues closely relate to the nature of neutrinos and  also play important roles in particle physics and cosmology \cite{review2}. 
A paper  \cite{neutrinoBAU} has shown that CPV in neutrino mixing is a necessary source of the generation of the baryon asymmetry of the Universe in popular leptogenesis scenarios. 
In addition, non-zero CPV must be a clear signal to guide new physics  beyond the Standard Model. 
Due to the more interesting implications and roles in phenomenology, Dirac CVP often garners greater attention. It exists for both Dirac neutrinos and Majorana fermions. Inspired by its similar form as CKM in the quark sector, 
it is also a bridge to establishing the relation between CPV in CKM and in the PMNS matrix.
Although Dirac CPV is essentially unknown, we still have some clues and hints from global fits of neutrino experiment data and flavor physics.

The updated neutrino oscillation data have been fitted for the CP phase by $\nu_e$ appearance data from long-baseline experiments in combination with precise $\theta_{13}$ to improve the sensitivity to  $\delta_{CP}$ \cite{NeuFit2014,Capozzi2014,Capozzi2016,Estaban2017}. The fit results show a maximal Dirac CPV, i.e., the center value of $3\pi/2$ for both NH and IN. 
The maximal CPV was first reported by the NO$\nu$A neutrino oscillation experiment \cite{MCPVinNOnuA}.
T2K also declared $\delta_{CP}=-\pi/2$ as the first hint \cite{Ghosh2017}.
On the other hand, flavor symmetry also predicts a maximal value of the phase such as A4 \cite{predictCPV,Petcov2015}. 
In the paper, we will focus on its indication in the neutrino mass hierarchy problem, especially at its most likely value: $3\pi/2$. 
The effect of leptonic CPV can be enhanced in neutrino oscillations with matter effects \cite{mattereffect}. 
When a neutrino travels through Earth matter, the asymmetric contributions to neutrino oscillations and anti-neutrino oscillations can break the asymmetry $A_{CP}^{\mu e}$ into two bands for a normal hierarchy (NH) and for an inverted hierarchy (IH). Due to the suppressed deviation at maximal Dirac CPV, the error bands of NH and IH decrease to form a gap. At the range of the gap between NH and IH, the neutrino mass hierarchy problem can be addressed.
The two principal issues concerning the nature of neutrinos, leptonic CPV and mass hierarchy, are connected. 
After a short review of  neutrino mixing, we will focus on the neutrino CP violation observable, the asymmetry $A_{CP}$, which helps to address the mass hierarchy problem. 
The ability of current long-baseline neutrino experiments to address the neutrino mass hierarchy is also investigated. The optimized neutrino detector performance, detector distance $L$ and energy will be  presented.
As the main result, the parameter space for detecting distance and neutrino energy is given.

The relation between the weak gauge eigenstates $\nu_{\alpha}$ ($\alpha=e,\mu,\tau$) and the mass eigenstates $\nu_{i}$  ($i=1,2,3$) can be expressed by a unitary rotation
\begin{equation}
|{\nu_\alpha}\rangle = \sum^3_{i=1}U^*_{\alpha i} |{\nu_i}\rangle
\end{equation}
with the $3 \times 3$ PMNS matrix $U$ 
\begin{eqnarray*} \label{Umatrix}
U&=&
  \begin{pmatrix}
   1 & 0 & 0 \\
    0 &c_{23}&s_{23} \\
     0& -s_{23} & c_{23} \\
    \end{pmatrix} 
    \begin{pmatrix}
    c_{13} & 0 & s_{13}e^{-i\delta} \\
    0& 1 &0 \\
     -s_{13}e^{i\delta}& 0&c_{13} \\
    \end{pmatrix} 
    \begin{pmatrix}
    c_{12}& s_{12} & 0 \\
    -s_{12} &c_{12}&0 \\
     0&0&1 \\
    \end{pmatrix}
     \text{diag}(e^{i\frac{\alpha_{1}}{2}},e^{i\frac{\alpha_{2}}{2}},1).
\\
    &=& \begin{pmatrix}
    c_{12}c_{13} & s_{12}c_{13} & s_{13}e^{-i\delta} \\
    -s_{12}c_{23}-c_{12}s_{23}s_{13}e^{i\delta} &c_{12}c_{23}-s_{12}s_{23}s_{13}e^{i\delta}&s_{23}c_{13} \\
     s_{12}s_{23}-c_{12}c_{23}s_{13}e^{i\delta}& -c_{12}s_{23}-s_{12}c_{23}s_{13}e^{i\delta}& c_{23}c_{13} \\
    \end{pmatrix} 
 \text{diag}(e^{i\frac{\alpha_{1}}{2}},e^{i\frac{\alpha_{2}}{2}},1).
 \end{eqnarray*}
The oscillation probability from flavor $\nu_\alpha$ to flavor $\nu_\beta$ at a distance of $L$ with energy $E$ is
\begin{eqnarray*}
P(\nu_\alpha\rightarrow\nu_\beta)=\Big|\sum_i U^*_{\alpha i}U_{\beta i}e^{-im_i^2L/(2E)}\Big|^2
\end{eqnarray*}
After expanding the above expression in terms of the small parameter $\alpha=\Delta m^2_{21}/\Delta m^2_{31}$, the CPV effect appears at the next-leading-order term and only contributes slightly.
A more convenient observable is an asymmetry between $\nu_\mu-\nu_e$  oscillations and $\bar{\nu}_{\mu}-\bar{\nu}_e$  oscillations
	\begin{eqnarray*}
		A_{CP}^{\mu e}&=&P(\nu_\mu\rightarrow \nu_{e})-P(\bar{\nu}_\mu-\bar{\nu}_{e})
		\\
		&=&-J_{CP}\Big\{\sin(\frac{\Delta m^2_{21}L}{2E})+\sin(\frac{\Delta m^2_{32}L}{2E})+\sin(\frac{\Delta m^2_{13}L}{2E})\Big\}
	\end{eqnarray*}
with $$J_{CP}=\frac{1}{8}\cos\theta_{13}\sin2\theta_{12}\sin2\theta_{23}\sin2\theta_{13}\sin\delta_{CP}.$$	
Due to the dependence on not only the absolute value of $\Delta m^2_{13}$ but also its  sign, $A_{CP}^{\mu e}$ provides a way to  address the neutrino mass hierarchy problem. 
Using the error transfer formula, the errors from the mixing angles, squared-mass difference, CP phase, and energy generate error bands for $A_{CP}^{\mu e}$. 
If there is a gap of $A_{CP}^{\mu e}$ between NH and IH, the neutrino mass hierarchy will be determined.
An excitation gap appears in neutrino oscillations with matter effects. 
When a neutrino travels through the Earth, the electrons in matter will interact with the neutrino. Due of a lack of anti-matter interactions, asymmetric contributions  from neutrino oscillations from anti-neutrino oscillations cause a separation of $A_{CP}$ for NH and IH \cite{mattereffect}.
The oscillation probability up to second order in the small parameters $\alpha$ and $s_{13}^2$ is approximated as 
\begin{eqnarray*}
	&&P^m(\nu_\mu\rightarrow\nu_e)=s_{23}^2\frac{\sin^2(2\theta_{13})}{(A-1)^2}\sin^2[(A-1)\Delta]
	\\
	&&-\frac{8\alpha}{A(1-A)}J_{CP}\Big\{\sin(\Delta)-\cos(\Delta)\Big\}\sin(A\Delta)\sin[\Delta(1-A)]
	\\
	&&+\alpha^2\cos^\theta_{23}\frac{\sin^2(2\theta_{12})}{A^2}\sin^2(A\Delta)
\end{eqnarray*}
where $A=\sqrt{2}2G_FN_eE/\Delta m_{31}^2$ with the electron number density of the Earth $N_e$
(for antineutrino oscillations, the probability can be obtained by changing $A\rightarrow -A$ and $\delta_{CP}\rightarrow-\delta_{CP}$).
In contrast to vacuum oscillations, the asymmetry  $A_{CP}^{\mu e}$ with matter effects depends on not only  the ratio of $E/L$ but also the distance or the neutrino energy itself.
\begin{figure}[htb]
\centering
\subfloat[L=50km]{%
  \includegraphics[width=.5\textwidth]{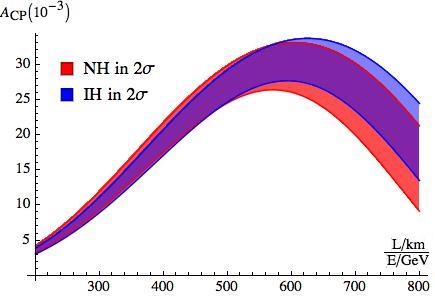}}\hfill
\subfloat[L=500km]{%
  \includegraphics[width=.5\textwidth]{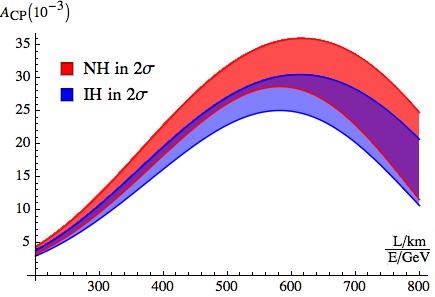}}\hfill
\\
\subfloat[L=1000km]{%
  \includegraphics[width=.5\textwidth]{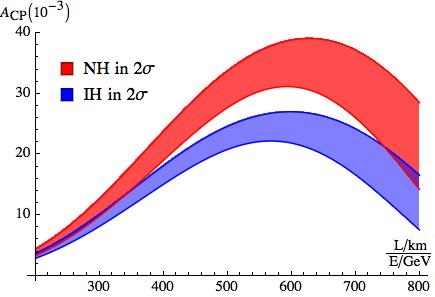}}\hfill
\subfloat[L=2000km]{%
  \includegraphics[width=.5\textwidth]{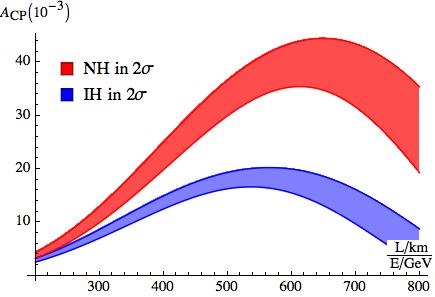}}
\caption{A gap of $A_{CP}$ for NH and IH with different lengths of  baselines in $2\sigma$ C.L.. The purple area is the overlay area from NH and IH. The best fit values of the mixing angles and squared-mass difference can be found in \cite{Capozzi2014}.}
\label{fig:gap}
\end{figure}
 Fig.\ref{fig:gap} shows that a gap appears with increasing source-detector distance $L$ due to the accumulation of asymmetric contributions from interactions with Earth matter. 
More precisely, the gap can be described by the number of statistical deviations $\lambda$,  defined as
	\begin{eqnarray}
		\lambda =\frac{\Big|A_{CP}^{\mu e}|_{NH}-A_{CP}^{\mu e}|_{IH}\Big|}{\delta(A_{CP}^{\mu e})|_{NH}+\delta(A_{CP}^{\mu e})|_{IH}}.
		\label{eq:lambda}
	\end{eqnarray}
The gap that appears corresponds to $\lambda=1$, i.e., NH and IH can be divided as $1\sigma$ C.L..

\begin{figure}[htb]
  \centering
 \includegraphics[width=0.5\textwidth]{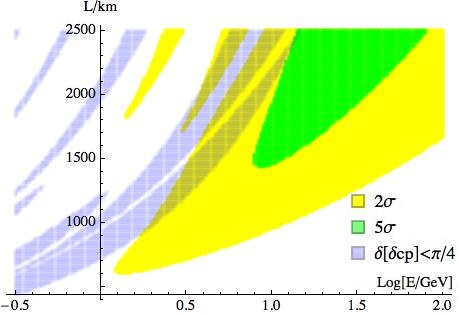}
\caption{The parameter space in the plane of $E$-$L$. The yellow area (the green area) indicates the separation of NH and IH in $2\sigma$ ($5\sigma$) C.L.. The purple area is allowed by the requirement of the standard deviation of Dirac CPV being less than $\pi/4$.}
\label{fig:paraspace}
\end{figure}

A general parameter space of $E$ and $L$ is shown in Fig.\ref{fig:paraspace}.
An extra requirement in the determination of Dirac CPV has been included.
Experiments have hinted at $\delta_{CP}=3\pi/2$ as a typical valve and have shown a large standard deviation simultaneously. A minimal limit of this standard deviation of less than $\pi/4$ is recommended \cite{Zhang2015}, which  corresponds to 95\% C.L. on a half cycle.
In the overlapping areas in Fig.\ref{fig:paraspace}, long-baseline neutrino oscillation experiment can determine the most likely Dirac CPV and address the mass hierarchy problem simultaneously. 
The minimal distance of the baseline is approximately $937$ km ($2400$ km) in $2\sigma$ ($5\sigma$) C.L. corresponding to a minimal beam energy of approximately $1.8 GeV$ ($13.3 GeV$). When ignoring the statistical requirement of the standard deviation of $\delta_{CP}$ or for a measured Dirac CPV, the minimal $L$ is $613$ km ($1437$ km) in $2\sigma$ ($5\sigma$) C.L. corresponding to a minimal $E$ of approximately $1.2 GeV$ ($7.9 GeV$).

Some long-baseline neutrino oscillation experiments have been run or planned to address the leptonic CVP and/or neutrino mass hierarchy problem.
Tab. \ref{tab:lambda} lists the $\lambda$ values of some long-baseline neutrino oscillation experiments at $\delta_{CP}=3\pi/2$ and the range of Dirac CPV corresponding to $1\sigma$ C.L.
\begin{table}[!hbp]
\centering
\begin{tabular}{|c|c|c|}
\hline
\hline
exp.&$\lambda|_{\delta_{CP}=3\pi/2}$ &$\delta_{CP}$ range for $\lambda>1$ 
\\
\hline
MINOS & $1.7$ & $[0.46,0.54]\pi$,$[1.46,1.54]\pi$
\\
\hline
NO$\nu$A & $2.6$ & $[0.45,0.55]\pi$,$[1.45,1.55]\pi$
\\
\hline
DUNE & $3.5$ & $[0.41,0.59]\pi$,$[1.41,1.59]\pi$
\\
\hline\hline
\end{tabular}
\caption{$\lambda$ values of current long-baseline neutrino oscillation experiments}
\label{tab:lambda}
\end{table} 
Here, the actual energy distinguishability has been considered in the table, which tends to increase the error in the denominator of Eq. \ref{eq:lambda} and suppress $\lambda$. Due to the sensitive energy error in high-energy neutrino beams, the suppressing effect on $\lambda$ will become more notable for high energy ranges. 
This will be an effective method for enhancing the confidence level by improving the energy resolution of detectors.
The data in Tab.\ref{tab:lambda} show that only a narrow window is open at maximal Dirac CPV.
The reason comes from  the error of the Dirac CPV.
Only at the point of maximal CPV does the contribution from the  energy resolution vanish. 
When considering Dirac CPV and taking the minimal distinguishable error, $\delta[\delta_{CP}]=\pi/4$, 
a relatively large error dominates the error band of $A_{CP}$. 
Thus, the window rapidly closes at a range slightly away from $\delta_{CP}=3\pi/2$. 

In conclusion, the asymmetric contributions to $A_{CP}^{\mu e}$ between NH and IH are enhanced  when neutrinos travel through Earth matter. 
Analyzing  the role of Dirac CPV can open a window to provide insight into the neutrino mass hierarchy problem at maximum  $\delta_{CP}$. A connection between two unknown essential issues characterizing the nature of neutrinos, Dirac CPV and mass hierarchy, has been established.
The minimal requirements on detecting distance and energy are $L>613$ km  and $E>1.2$ GeV in $2\sigma$ C.L. (without the distinguishability requirement).  

\section*{Acknowledgement}
The author would like to thank R. Li for useful discussions. 
This work was supported in part also by the Fundamental Research Funds for the Central Universities and by NSFC Grant 11775165.


\begin{thebibliography}{1}
\bibitem{review1}
	R.N. Mohapatra et al, Rept.Prog.Phys. 70 (2007) 1757;
	J.W.F. Valle, J.Phys.Conf.Ser. 53 (2006) 473.  
\bibitem{review2}
	G.C. Branco, R.Gonzalez Felipe, F.R. Joaquim, Rev.Mod.Phys. 84 (2012) 515.
\bibitem{neutrinoBAU}
	S.T. Petcov, Nucl.Phys. B908 (2016) 279.
\bibitem{NeuFit2014}
	D.V. Forero,  M. T\'{o}rtola, and J.W.F. Valle, Phys. Rev. D 90, 093006 (2014).
\bibitem{Capozzi2014}
	F. Capozzi, et al, Phys. Rev. D 89, 093018 (2014).
\bibitem{Capozzi2016}
	F. Capozzi, E. Lisi, A. Marrone, D. Montanino, A. Palazzo, Nucl. Phys. B908 (2016) 218.
\bibitem{Estaban2017}
	I. Esteban, M.C. Gonzalez-Garcia, M. Maltoni, I. Martinez-Soler, T. Schwetz, JHEP 1701 (2017) 087.
\bibitem{MCPVinNOnuA}
	P. Adamson, et al., Phys. Rev. Lett. 116, 151806 (2016);
	F. Capozzi, et al., Nucl. Phys. B 908, 218 (2016). 
\bibitem{Ghosh2017}
	M. Ghosh, arXiv:1702.0788.
\bibitem{predictCPV}
	I. Girardi, S.T. Petcov, A.V. Titov, Int.J.Mod.Phys. A30 (2015) 1530035.
(2015).
\bibitem{Petcov2015}
	S.T. Petcov, Nucl. Phys. B 892 (2015) 400.
\bibitem{mattereffect}
	R. Gandhi, P. Ghoshal, S. Goswami, P. Mehta, S.U. Sankar, Phys.Rev. D73 (2006) 053001.
\bibitem{Zhang2015}
	K. Fu, Y. Zhang, Phys. Lett. B 746 (2015) 104.
\end{thebibliography}
\end{document}